\documentclass[a4paper,11pt]{article}
\pdfoutput=1
\usepackage{jheparxiv}
\usepackage[T1]{fontenc}

\usepackage{graphicx,epsfig,latexsym,amsfonts,amsmath,amsthm,amssymb,amsbsy,multirow,slashed,wasysym,textcomp,subfigure,wrapfig,datetime,comment,mathtools,cancel,mathrsfs,tikz-cd,mathtools,bbm,enumitem,xcolor,twistor}

\usepackage{soul}

\renewcommand{\d}{\mathrm{d}}

\newcommand{\da}{{\dot\alpha}}
\newcommand{\nbar}{{\overline\nabla}}

\newcommand{\Dbar}{{\overline{\mathrm{D}}}}

\title{\boldmath Schwarzschild black holes from twistor space}

\author[a]{Tim Adamo,}
\author[a]{Bernardo Araneda,}
\author[a]{Sean Seet}
\author[b]{\& Atul Sharma}

\affiliation[a]{School of Mathematics and Maxwell Institute for Mathematical Sciences,\\
University of Edinburgh, EH9 3FD, UK\vspace{0.1cm}}
\affiliation[b]{Center for the Fundamental Laws of Nature \& Black Hole Initiative,\\Harvard University, Cambridge, MA, 02138, USA \vspace{0.1cm}}

\emailAdd{t.adamo@ed.ac.uk}
\emailAdd{baraneda@ed.ac.uk}
\emailAdd{sseet@ed.ac.uk}
\emailAdd{atulsharma@fas.harvard.edu}

\abstract{Twistor theory forms the basis for many surprising advances in areas ranging from dynamical systems to quantum field theory. Yet for almost fifty years, one of the main drawbacks of twistor theory has been its inability to give non-perturbative descriptions of non-chiral (or non-self-dual) field configurations. This difficulty is known as `the googly problem.' In this paper, we provide a resolution of the googly problem for a particular solution of the vacuum Einstein equations: the Schwarzschild metric. We start with the twistor space of the self-dual Taub-NUT Euclidean gravitational instanton, expressed in Kerr-Schild form. Within this twistor space, we then consider a quadric which corresponds to the anti-self-dual Taub-NUT metric. While the full quadric is not holomorphic with respect to the complex structure of the self-dual Taub-NUT twistor space, its holomorphic locus still has complex dimension two. This `coincidence locus' -- points in twistor space on the holomorphic portion of the quadric -- inherits a complex structure from the twistor space and a symplectic form from the quadric itself. Remarkably, these structures are compatible, giving rise to a non-self-dual, four-dimensional K\"ahler metric which is conformal to Schwarzschild (in Lorentzian or Euclidean signature). This is the first instance of a non-self-dual Einstein metric constructed entirely from holomorphic data in a twistor space.}

\begin{document}
\maketitle
\flushbottom

\section{Introduction}

Twistor theory provides a remarkable route to generating solutions of non-linear partial differential equations (PDEs)~\cite{Penrose:1967wn,Penrose:1972ia}. Broadly speaking, this proceeds by taking holomorphic data on a certain complex variety -- a \emph{twistor space} -- and producing field configurations which obey the desired equations. Crucially, the equations are solved \emph{automatically}: holomorphicity of the twistor data ensures that the PDEs are solved. In other words, twistor theory solves non-linear PDEs with free analytic data.

While the first application of twistor theory was to linear PDEs -- the zero-rest-mass equations of arbitrary spin in four-dimensional flat space~\cite{Penrose:1969ae} -- it was soon realised that solutions to highly non-linear equations could also be generated in this way, including the self-dual Yang-Mills~\cite{Ward:1977ta} and self-dual Einstein equations~\cite{Penrose:1976js}. Furthermore, \emph{all} solutions of these equations could be generated from twistor data.

Alas, the self-dual Einstein equations are only a subsector of the full Einstein equations, and -- as there are no non-trivial Lorentzian-real self-dual metrics -- apparently not a physically relevant one. Furthermore, the self-duality equations, while non-linear, are integrable~\cite{Belavin:1978pa,Boyer:1985aj}, a fact which is immediately established by their twistor construction, translating these PDEs into the integrability condition of a complex structure or partial connection. Indeed, by the 1990s it was clear that all of the non-linear PDEs captured by twistor theory were integrable, being symmetry reductions or non-commutative deformations of the self-duality equations~\cite{Ward:1985gz,Mason:1991rf} -- see~\cite{Hitchin:1983ay,Hitchin:1986vp,Mason:1989kk,Mason:1992vd,Strachan:1992em} for some well-studied examples.

This difficulty -- the apparent inability to encode solutions to non-self-dual or non-integrable equations -- in twistor space was termed\footnote{This (somewhat culturally specific) name refers to a particular delivery in cricket, in which the bowler spins the ball in the opposite direction to usual in order to deceive the batter.} `the googly problem'~\cite{Penrose:1968me,Penrose:1999cw,Penrose:1999sz,Penrose:2015lla}. For a theory whose initial ambition was to provide a route to quantum gravity, this seemed a major obstacle, and significant effort over the years has been devoted towards overcoming it -- without success (though there continue to be intriguing proposals~\cite{Penrose:2015lla,Marcolli:2020zfc}).

Yet despite the persistence of the googly problem, twistor theory has proven a powerful and widely applicable tool. The googly problem can be solved \emph{perturbatively}~\cite{Witten:2003nn,Mason:2005zm,Skinner:2013xp,Sharma:2021pkl} thanks to reformulations of Yang-Mills~\cite{Chalmers:1996rq} and general relativity~\cite{Plebanski:1977zz} which organize perturbation theory around the self-dual sector. This has enabled all multiplicity calculations of gluon and graviton scattering amplitudes~\cite{Roiban:2004yf,Cachazo:2012kg}; proofs of relationships between amplitudes, Wilson loops and correlation functions to all-loop orders in planar $\cN=4$ super-Yang-Mills theory~\cite{Mason:2010yk,Bullimore:2011ni,Adamo:2011dq,Adamo:2011cd,Koster:2016ebi}; bootstrap calculations of high-precision scattering in massless QCD~\cite{Costello:2023vyy,Dixon:2024mzh,Dixon:2024tsb,Morales:2025alm}; and even a new method for determining the 1-loop QCD $\beta$-function~\cite{Bittleston:2025jmk}. Even within the self-dual sector, twistor theory has provided a route to holographic descriptions of integrable systems~\cite{Costello:2022jpg,Costello:2023hmi,Bittleston:2024efo}; illuminated the hyperk\"ahler geometry associated with Donaldson-Thomas invariants~\cite{Bridgeland:2020zjh,Bridgeland:2024ydg,Dunajski:2024hxj}; and underpinned methods of systematically obtaining lower-dimensional integrable models~\cite{Bittleston:2020hfv,Penna:2020uky,Cole:2023umd,Cole:2024sje,Bittleston:2026tdr,Ashwinkumar:2026zrr}.

\medskip

It then seems that twistor theory has managed to thrive over the last fifty years despite the googly problem remaining un-resolved at the fully non-linear level. Of course, the googly problem still presents an obstruction to interpreting twistor theory as a fundamental description of gravity or gauge theory (at the classical or quantum levels). It also rears its head in a more practical sense: recently, twistor methods have been used to compute graviton scattering amplitudes \emph{exactly} on curved, self-dual metrics~\cite{Adamo:2020syc,Adamo:2021bej,Adamo:2022mev}, including the self-dual Taub-NUT metric~\cite{Adamo:2025fqt}. While these calculations demonstrate the power of twistor theory -- they would be impossible with standard background field theory techniques -- one is left wondering what, if anything, such surprising results can tell us about gravitational scattering on real, Lorentzian spacetimes like astrophysical black holes. Addressing this surely requires some resolution of the googly problem at least in the limited context of a fixed background metric.

What properties should a twistor construction possess to qualify as a solution of the googly problem (even for a specific metric)? Motivated by the desire to preserve as many features of the twistor construction of self-dual non-linear fields as possible, it seems that three fairly obvious criteria emerge. First, the construction must use only data on a twistor space, rather than a combination of twistor and dual twistor spaces\footnote{That is not to say that a mixture of twistor and dual twistor data cannot lead to non-integrable field equations. This is precisely the setup of ambitwistor theory~\cite{Isenberg:1978kk,LeBrun:1983}, although field equations require the language of formal neighbourhoods~\cite{Witten:1985nt,Baston:1987av,LeBrun:1991jh} or combination with worldsheet methods~\cite{Mason:2013sva,Adamo:2014wea}.}. Secondly, this twistor data should only be subject to \emph{holomorphic} constraints, in keeping with the general theme that physical solutions are encoded by analytic data on twistor space. Finally, no field equations should be explicitly solved on twistor space -- the resulting metric should solve the Einstein equations automatically as a consequence of the twistor construction, as in the self-dual case.

\medskip

In this paper, we give a construction which satisfies all three of these criteria to provide a solution of the googly problem for the specific case of the Schwarzschild metric. Our construction is inspired by a combination of the facts that \emph{non}-self-dual black holes have conformally K\"ahler structures~\cite{Flaherty:1974,Flaherty:1976,Aksteiner:2022bwr}, and that self-dual K\"ahler structures are closely related to special surfaces in twistor space~\cite{Pontecorvo:1992}. A framework connecting non-self-dual black hole spacetimes and quadric surfaces in flat twistor space was developed in~\cite{Araneda:2022xii}. A twistor quadric defines both a symplectic structure and a complex structure, and the construction in~\cite{Araneda:2022xii} is based on fixing the former but deforming the latter, in such a way that the new structures are still compatible. These define, by composition, a new, non-self-dual K\"ahler metric on spacetime, recovering in particular the standard black hole solutions. 

It has, however, remained unclear how the deformed quadrics are related to new holomorphic structures in \emph{some} ambient twistor space, and similarly, how to find the deformations that lead \emph{automatically} to conformally Einstein metrics. In this paper we solve both of these issues in a unified way for the Schwarzschild metric, leading to a solution of the googly problem for this case.

Our construction is also motivated by two other ideas in the literature, one old and one more recent. The first is Hawking's observation that the Schwarzschild metric (in Euclidean signature) should be regarded as a combination of the self-dual and anti-self-dual Taub-NUT metrics~\cite{Hawking:1976jb}, in analogy to a Coulomb charge in electromagnetism arising from the superposition of self-dual and anti-self-dual dyons. This picture has been realized through an explicit superposition prescription~\cite{Kim:2024mpy}, making use of the recently discovered Kerr-Schild form of the (anti-)self-dual Taub-NUT metric~\cite{Kim:2024dxo}.

Further, it has been shown (in an effort to make a connection between~\cite{Araneda:2022xii} and the standard non-linear graviton) that the anti-self-dual Taub-NUT metric can be obtained using holomorphic quadrics in the `wrong' twistor space~\cite{Adamo:2026obu}. The standard way to obtain anti-self-dual Taub-NUT would be to implement the non-linear graviton construction~\cite{Penrose:1976js} in dual twistor space (i.e., the projective dual of the twistor space associated with self-duality), but in~\cite{Adamo:2026obu} it is constructed in Kerr-Schild form by placing a holomorphic quadric into the twistor space of flat space. This `googly quadric' procedure avoids the complex deformation that underlies the usual non-linear graviton construction, and is tied to the algebraic speciality of the Taub-NUT metric.  

Here, we combine all of these observations directly in twistor space. We first construct the twistor space $\CPT$ of self-dual Taub-NUT in Kerr-Schild gauge using the standard non-linear graviton construction. Then we consider the quadric corresponding to anti-self-dual Taub-NUT inside $\CPT$. In the complex structure of $\CPT$, this quadric is \emph{not} holomorphic, but one can ask for the subvariety of the quadric which is holomorphic. We call this subvariety $\chi$ the \emph{coincidence locus}, as it is specified by the coincidence of two equations: the quadric constraint and the holomorphicity condition. 

Na\"ively, these two equations in three (complex) dimensional twistor space will define a variety of one complex dimension. Surprisingly, one finds that $\dim_{\C}\chi=2$ thanks to the subtle interplay between the self-dual and anti-self-dual Taub-NUT geometric structures. The complex structure of $\CPT$ and the symplectic structure naturally associated to the quadric then induce a K\"ahler metric on $\chi$ which is conformal to the Schwarzschild metric. This fact -- stated as Theorem~\ref{MainThm} below -- is the central result of this paper.

This construction satisfies all three of our criteria for a solution of the googly problem: it is set entirely in twistor space (namely, the twistor space of self-dual Taub-NUT); it involves only holomorphic data (namely, the coincidence locus); and it does not involve explicit solution of the Einstein equations (the Schwarzschild metric emerges automatically). Furthermore, while the inputs of the construction are the Euclidean self-dual and anti-self-dual Taub-NUT metrics, the resulting Schwarzschild metric can be given either Euclidean or Lorentzian reality conditions. In particular, this means that the true Schwarzschild black hole metric is obtained in this way.

\medskip

The paper is organized as follows. Section~\ref{sec:TwistBack} details the twistor construction of self-dual Taub-NUT in Kerr-Schild gauge, as well as the construction of anti-self-dual Taub-NUT using a twistor quadric. Section~\ref{sec:Construction} then defines the coincidence locus, its resulting complex and K\"ahler structures and the fact that they encode the Schwarzschild metric. Section~\ref{sec:Disc} then discusses future directions for research building on this work, including obtaining other solutions and potential applications in black hole perturbation theory.


\section{(A)SD Taub-NUT and twistor space}\label{sec:TwistBack}

The standard way in which vacuum self-dual (SD) or anti-self-dual (ASD) metrics are realised with twistor theory is through the non-linear graviton construction~\cite{Penrose:1976js}. In essence, one begins with either the twistor or dual twistor space of complexified Minkowski spacetime: SD metrics correspond to integrable complex structure deformations of twistor space while ASD metrics correspond to such deformations of dual twistor space. In this section, we discuss how two particular metrics -- the SD and ASD Taub-NUT metrics -- can \emph{both} be obtained from structures in twistor space. While the non-linear graviton construction of these metrics (in twistor or dual twistor space) is well-known~\cite{Sparling:1976,Hitchin:1979rts,lebrun1991complete,Adamo:2025fqt}, the constructions utilized here are non-standard in both instances. 

In particular, for the SD Taub-NUT (SDTN) metric we make use of a non-linear graviton construction which gives rise to the metric in Kerr-Schild form. As this Kerr-Schild form of the SDTN metric was only discovered two years ago~\cite{Kim:2024dxo}, the corresponding non-linear graviton construction has not yet appeared in the literature. For the ASD Taub-NUT (ASDTN) metric, we make use of a recent construction for (complex) hyperk\"ahler metrics of type D -- the `self-dual black hole' metrics -- from quadric hypersurfaces in twistor (rather than dual) twistor space~\cite{Adamo:2026obu}. 


\subsection{Self-dual Taub-NUT in Kerr-Schild form}

The self-dual Taub-NUT metric (SDTN), famously a hyperk\"ahler, asymptotically locally flat gravitational instanton in Euclidean signature~\cite{Hawking:1976jb,Gibbons:1978tef,Gibbons:1979xm}, is typically presented in Gibbons-Hawking coordinates $(\tau,R,\theta,\phi)$:
\be\label{SDTN-GH}
\d s^2=V^{-1}\left(\d\tau+A\right)^2+V\left(\d R^2+R^2\,\d\Omega^2\right)\,,
\ee
where $\d\Omega^2$ is the round line element on $S^2$ with coordinates $(\theta,\phi)$ and
\be\label{monopole}
V=1+\frac{2\,M}{R}\,, \qquad \d A=\star_{3}\d V\,,
\ee
with $\star_3$ the Hodge star on $\R^3$ with coordinates $(R,\theta,\phi)$ and the flat metric $\d s^2_{\R^3}=\d R^2+R^2\,\d\Omega^2$. This presentation makes self-duality manifest through the monopole equation satisfied by the scalar potential $V$ and the 1-form $A$. All of the twistor descriptions of this metric in the literature use the non-linear graviton construction to obtain the SDTN metric in this (or a closely related) Gibbons-Hawking form. 

Very recently, it was realised that the SDTN metric also admits a Kerr-Schild coordinate system \cite{Kim:2024dxo}\footnote{It should also be noted that this Kerr-Schild form of SDTN can be used to justify its interpretation as a `self-dual black hole'~\cite{Desai:2024fgr} and has an interesting interplay with the Penrose transform~\cite{Albertini:2025ogf}.}. In particular, this means that the (fully non-linear) metric can be described as a finite perturbation of the flat metric:
\be\label{SDTN-KS}
\d s^{2}=\d \tau^2+\d x^2+\d y^2+\d z^2-\frac{2M}{r}\left[\d\tau-\im\,\d z+\left(\frac{z+r}{x+\im\,y}\right)\left(\im\,\d x-\d y\right)\right]^2 \,,
\ee
where $r^2\equiv x^2+y^2+z^2$. The diffeomorphism between these Kerr-Schild coordinates and the well-known Gibbons-Hawking coordinates of \eqref{SDTN-GH} is known explicitly~\cite{Adamo:2026obu}. The Kerr-Schild form \eqref{SDTN-KS} can be expressed in an illuminating way after making some identifications with natural 2-spinor quantities associated with the SDTN metric. 

In the 2-spinor formalism, we raise and lower $\mathfrak{sl}(2,\C)$ spinors with the 2-dimensional Levi-Civita symbols, following the conventions of~\cite{Adamo:2017qyl}:
\be\label{2-spinors}
\epsilon^{\alpha\beta}\,a_{\beta}\,b_{\alpha}=a^{\alpha}\,b_{\alpha}\equiv \la a\,b\ra\,, \qquad \epsilon^{\dot\alpha\dot\beta}\,\tilde{a}_{\dot\beta}\,\tilde{b}_{\dot\alpha}=\tilde{a}^{\dot\alpha}\,\tilde{b}_{\dot\alpha}\equiv [\tilde{a}\,\tilde{b}]\,.
\ee
It will also be convenient to introduce a normalised spinor dyad $\{o_{\alpha},\iota_{\alpha}\}$, $\la\iota\,o\ra=1$, and similarly for the dotted spinors. An explicit realization of this dyad is simply given by $o_{\alpha}=(1,0)$ and $\iota_{\alpha}=(0,1)$. 

Now, the coordinates $(\tau,x,y,z)$ can be packaged into the $2\times 2$ matrix
\be\label{flatcoords}
x^{\alpha\dot\alpha}=\frac{1}{\sqrt{2}}\left(\begin{array}{cc}
                                              \tau-\im\,z & \im\,x-y \\
                                              \im\,x+y & \tau+\im z
                                              \end{array}\right)\,,
\ee
so that the flat metric is written as
\be\label{flatmet}
\delta=\epsilon_{\alpha\beta}\,\epsilon_{\dot\alpha\dot\beta}\,\d x^{\alpha\dot\alpha}\,\d x^{\beta\dot\beta}\,.
\ee
The metric \eqref{SDTN-KS} has a tri-holomorphic Killing vector $T=\partial_{\tau}$, which is written in the 2-spinor formalism as
\be\label{Tvec}
T^{\alpha\dot\alpha}=\frac{1}{\sqrt{2}}\left(\begin{array}{cc}
                                              1 & 0 \\
                                              0 & 1
                                              \end{array}\right)=\frac{1}{\sqrt{2}}\left(o^{\alpha}\,o^{\dot\alpha}+\iota^{\alpha}\,\iota^{\dot\alpha}\right)\,.
\ee
Furthermore, the SDTN metric is algebraically special of type D and admits a self-dual Killing spinor $K^{\dot\alpha\dot\beta}=K^{\dot\beta\dot\alpha}$ which obeys
\be\label{SDTN-Killing}
\nabla^{\gamma(\dot\gamma}K^{\dot\alpha\dot\beta)}=0\,, \qquad \frac{2}{3}\,\nabla^{\alpha}{}_{\dot\beta}K^{\dot\alpha\dot\beta}=T^{\alpha\dot\alpha}\,.
\ee
This Killing spinor can be decomposed as
\be\label{SDTN-pSpinors}
K^{\dot\alpha\dot\beta}=\alpha^{(\dot\alpha}\,\beta^{\dot\beta)}\,,
\ee
with
\be\label{alphabeta}
\alpha^{\dot\alpha}=\sqrt{\frac{y-\im\,x}{2}}\left(o^{\dot\alpha}+\zeta_-\,\iota^{\dot\alpha}\right)\,, \qquad \beta^{\dot\alpha}=\sqrt{\frac{y-\im\,x}{2}}\left(o^{\dot\alpha}+\zeta_+\,\iota^{\dot\alpha}\right)\,,
\ee
and
\be\label{zetas}
\zeta_{\pm}:=\frac{z\mp r}{x+\im\,y}\,.
\ee
The spinors $\alpha^{\dot\alpha}$, $\beta^{\dot\beta}$ will often be referred to as the principal spinors of SDTN (in the Kerr-Schild coordinate system).

These quantities obey some elementary identities which will prove useful:
\be\label{spinident1}
[\beta\,\alpha]=\im\,r\,, \qquad \zeta_+\,\bar{\zeta}_-=-1\,.
\ee
These can be used to re-write the metric \eqref{SDTN-KS} as
\be\label{SDTN-KS2}
\begin{split}
\d s^2&=\delta-2\im\,M\,\frac{o_{\alpha}\,o_{\beta}\,\alpha_{\dot\alpha}\,\alpha_{\dot\beta}}{[\beta\,\alpha]\,\la o|T|\alpha]^2}\,\d x^{\alpha\dot\alpha}\,\d x^{\beta\dot\beta} \\
&=\delta-\frac{8\,M}{r\,(y-\im\,x)}\,\left(o_{\alpha}\,\alpha_{\dot\alpha}\,\d x^{\alpha\dot\alpha}\right)^{2}\,,
\end{split}
\ee
where the fact that the metric is Kerr-Schild is now very explicit, with
\be\label{KS-vec}
k^{\alpha\dot\alpha}:=o^{\alpha}\,\alpha^{\dot\alpha}\,,
\ee
identified as the Kerr-Schild null vector. This data is conveniently packaged into a self-dual Maxwell field
\be\label{SDMax}
\varphi_{\dot\alpha\dot\beta}:=-\frac{4M}{r}\,(o_{\dot\alpha}+\zeta_{-}\,\iota_{\dot\alpha})\,(o_{\dot\beta}+\zeta_{-}\,\iota_{\dot\beta})\,,
\ee
which obeys
\be\label{SDMax2}
\nabla^{\alpha\dot\alpha}\varphi_{\dot\alpha\dot\beta}=0=\varphi^{\dot\alpha\dot\beta}\,\varphi_{\dot\alpha\dot\beta}\,.
\ee
The SDTN metric in Kerr-Schild gauge can again be re-expressed as
\be\label{SDTN-KS3}
\d s^{2}=\delta+o_{\alpha}\,o_{\beta}\,\varphi_{\dot\alpha\dot\beta}\,\d x^{\alpha\dot\alpha}\,\d x^{\beta\dot\beta}\,,
\ee
in terms of this null, SD Maxwell field.

\medskip

As the SDTN metric is vacuum and self-dual -- that is, hyperk\"ahler -- it must admit a description in twistor space via Penrose's non-linear graviton construction~\cite{Penrose:1976js}. While the Gibbons-Hawking form of SDTN has been described with twistor theory many times -- indeed, it was one of the first explicit examples of the non-linear graviton~\cite{Sparling:1976,Hitchin:1979rts} -- the Kerr-Schild form of SDTN has not yet been given a twistorial treatment in the literature. Of course, the non-linear graviton construction for SDTN in Kerr-Schild form is neither difficult nor non-standard, but we will present it here in some detail for completeness.

As SDTN is a smooth, complete metric on $\R^4$, we will work with the Atiyah-Hitchin-Singer form of the non-linear graviton construction~\cite{Atiyah:1978wi}, which is naturally adapted to Riemannian hyperk\"ahler metrics. The twistor space of $\R^4$ with the flat metric \eqref{flatmet} is given by $\PT\cong\R^4\times\P^1$, with coordinates $(x^{\alpha\dot\alpha},[\lambda_{\alpha}])$, where $[\lambda_{\alpha}]$ denotes the holomorphic homogeneous coordinates $\lambda_{\alpha}=(\lambda_0,\lambda_1)$ on $\P^1$ considered up to projective rescalings $\lambda_\alpha\sim b\,\lambda_{\alpha}$ for any non-zero complex number $b$. The notion of complex conjugation which is compatible with Euclidean reality conditions acts on 2-spinors as
\be\label{hat}
\lambda_{\alpha}\mapsto\hat{\lambda}_{\alpha}=(-\bar{\lambda}_{1},\,\bar{\lambda}_0)\,,
\ee
which is simply the antipodal map on the Riemann sphere $\P^1$.

The flat twistor space $\PT$ is equipped with its natural complex structure, encoded in the integrable Dolbeault operator~\cite{Woodhouse:1985id}
\be\label{flatdbar}
\begin{split}
\dbar &=\frac{\D\hat{\lambda}}{\la\lambda\,\hat{\lambda}\ra}\,\lambda^{\alpha}\,\frac{\partial}{\partial\hat{\lambda}_{\alpha}}+\frac{\hat{\lambda}_{\alpha}\,\d x^{\alpha\dot\alpha}}{\la\lambda\,\hat{\lambda}\ra}\,\lambda^{\beta}\,\frac{\partial}{\partial x^{\beta\dot\alpha}} \\
 &=\bar{e}^{0}\,\dbar_0+\bar{e}^{\dot\alpha}\,\dbar_{\dot\alpha}\,,
\end{split}
\ee
where $\D\hat{\lambda}:=\la\hat{\lambda}\,\d\hat{\lambda}\ra$ and
\be\label{flat(0,1)basis}
\bar{e}^0:=\frac{\D\hat{\lambda}}{\la\lambda\,\hat{\lambda}\ra^2}\,, \qquad \bar{e}^{\dot\alpha}:=\frac{\hat{\lambda}_{\alpha}\,\d x^{\alpha\dot\alpha}}{\la\lambda\,\hat{\lambda}\ra}\,, \qquad \dbar_0:=\la\lambda\,\hat{\lambda}\ra\,\lambda^{\alpha}\,\frac{\partial}{\partial\hat{\lambda}^{\alpha}}\,, \qquad \dbar_{\dot\alpha}:=\lambda^{\alpha}\,\frac{\partial}{\partial x^{\alpha\dot\alpha}}
\ee
form a basis of $(0,1)$-forms and $(0,1)$-vectors, respectively.

The non-linear graviton theorem states that SDTN corresponds to an integrable complex structure deformation of $\PT$ which preserves the holomorphic fibration over $\P^1$ as well as the fiberwise symplectic structure $\lambda_\al\lambda_\beta\d x^{\al\da}\wedge\d x^{\beta}{}_\da$. Now, consider the complex structure deformation defined by
\be\label{SDTN-Dolb}
\nbar=\dbar+\frac{\la o\,\lambda\ra^2}{\la\lambda\,\hat{\lambda}\ra^2}\,\varphi_{\dot\alpha}{}^{\dot\beta}\,\hat{\lambda}_{\alpha}\,\hat{\lambda}^{\beta}\,\d x^{\alpha\dot\alpha}\,\frac{\partial}{\partial x^{\beta\dot\beta}}\,.
\ee
This almost complex structure is easily seen to be integrable, $\nbar^2=0$, and fiberwise symplectic, thanks to the basic properties \eqref{SDMax2} of the SD Maxwell field $\varphi_{\dot\alpha\dot\beta}$. And it clearly preserves the holomorphic fibration over $\P^1$, as the deformation does not point in these directions.

Let $\CPT$ denote $\R^4\times\P^1$ with the complex structure defined by $\nbar$; by the non-linear graviton construction, $\CPT$ will correspond to some hyperk\"ahler metric on (a suitably convex open subset of) $\R^4$. To recover this metric, observe that
\be\label{SDTN(1,0)basis}
\theta^{0}=\D\lambda\vcentcolon=\la\lambda\,\d\lambda\ra\,, \qquad \theta^{\dot\alpha}=\lambda_{\alpha}\,\d x^{\alpha\dot\alpha}-\frac{\la o\,\lambda\ra^2}{\la\lambda\,\hat{\lambda}\ra}\,\varphi^{\dot\alpha}{}_{\dot\beta}\,\hat{\lambda}_{\beta}\,\d x^{\beta\dot\beta}\,,
\ee
give a basis of $(1,0)$-forms on $\CPT$ in the new complex structure. With these, one can construct the Gindikin 2-form~\cite{Gindikin:1986}
\be\label{Gind1}
\Sigma=\theta^{\dot\alpha}\wedge\theta_{\dot\alpha}=\lambda_{\alpha}\,\lambda_{\beta}\left(\d x^{\alpha\dot\alpha}\wedge\d x^{\beta}{}_{\dot\alpha}- o^{\alpha}\,o^{\beta}\,\varphi_{\dot\alpha\dot\beta}\,\d x^{\gamma\dot\alpha}\wedge\d x_{\gamma}{}^{\dot\beta}\right)\,,
\ee
which obeys $\d(\Sigma\wedge\theta^0)=0$. This implies that a hyperk\"ahler metric is encoded in $\Sigma$ via its tetrad $e^{\alpha\dot\alpha}$, which can be identified as $\Sigma=\lambda_{\alpha}\lambda_{\beta} e^{\alpha\dot\alpha}\wedge e^{\beta}{}_{\dot\alpha}$. One finds
\be\label{SDTN-tetrad}
e^{\alpha\dot\alpha}=\d x^{\alpha\dot\alpha}+\frac{1}{2}o^{\alpha}\,o_{\beta}\,\varphi^{\dot\alpha}{}_{\dot\beta}\,\d x^{\beta\dot\beta}\,,
\ee
which is the tetrad for the SDTN metric \eqref{SDTN-KS3}, as desired\footnote{Strictly speaking, this construction determines the tetrad only up to a local Lorentz rotation acting on the dotted SU$(2)$ index, and every tetrad in this orbit correctly reproduces the SDTN metric.}. 

To summarize, this establishes that $\CPT$ with complex structure given by $\nbar$ is the twistor space of the SDTN metric in Kerr-Schild form. It goes without saying that the other well-known twistor descriptions of SDTN in the literature~\cite{Sparling:1976,Hitchin:1979rts,lebrun1991complete,Adamo:2025fqt} will be related to this one by gauge transformations of the Dolbeault operator.


\subsection{Anti-self-dual Taub-NUT as a twistor quadric}

One can equally well consider the \emph{anti}-self-dual Taub-NUT metric (ASDTN), which can also be written in Kerr-Schild form:
\be\label{ASDTN}
\d s^2=\delta -4\im\,M\,\frac{A_{\alpha}\,A_{\beta}\,o_{\dot\alpha}\,o_{\dot\beta}}{\la B\,A\ra\,\la A|T|o]^2}\,\d x^{\alpha\dot\alpha}\,\d x^{\beta\dot\beta}\,,
\ee
where $A_{\alpha}$, $B_{\alpha}$ are the principal spinors of ASDTN, which are related to those of SDTN in a remarkably simple way:
\be\label{AB}
A_{\alpha}=\sqrt{2}\,T_{\alpha}{}^{\dot\alpha}\,\alpha_{\dot\alpha}\,, \qquad B_{\alpha}=\sqrt{2}\,T_{\alpha}{}^{\dot\alpha}\,\beta_{\dot\alpha}\,.
\ee
As a vacuum, anti-self-dual metric, ASDTN can be recovered by performing the non-linear graviton construction in \emph{dual} twistor space (i.e., the projective dual of $\PT$). This is the essence of twistor theory's googly problem: the non-linear graviton construction gives metrics with either SD or ASD curvature (depending on whether one works in twistor or dual twistor space), but not both.

However, there is another construction of ASDTN which is distinct from the usual non-linear graviton and is formulated entirely in $\PT$, rather than its dual~\cite{Adamo:2026obu}. This exploits the fact that ASDTN has special geometric structures beyond being (anti-)hyperk\"ahler -- it is also algebraically special of Petrov type D. This means that the metric can be recovered by considering a holomorphic quadric in $\PT$ corresponding to its ASD Killing spinor.

In particular, consider the quadric (i.e., quadratic variety) 
\be\label{quadric}
\mathcal{Q}=\left\{(x,[\lambda])\in\PT\,|\, Q:=\la A\,\lambda\ra\,\la B\,\lambda\ra=0\right\}\,,
\ee
inside of $\R^4\times \P^1$, equipped with its trivial complex structure \eqref{flatdbar}. Observe that, as $A_{\alpha}$, $B_{\alpha}$ are the principal spinors of ASDTN, 
\be\label{quadric2}
Q=\la A\,\lambda\ra\,\la B\,\lambda\ra=\tilde{K}^{\alpha\beta}\,\lambda_{\alpha}\,\lambda_{\beta}\,,
\ee
where $\tilde{K}^{\alpha\beta}$ is the Killing 2-spinor of ASDTN, obeying $\nabla^{(\gamma}{}_{\dot\gamma}\tilde K^{\alpha\beta)}=0$. This ensures that
\be\label{holoquad}
\dbar Q=-\bar{e}^{\dot\alpha}\,\lambda_{\alpha}\,\lambda_{\beta}\,\lambda_{\gamma}\,\nabla^{\gamma}{}_{\dot\alpha}\tilde{K}^{\alpha\beta}=0\,,
\ee
meaning that $\mathcal{Q}\subset\PT$ is a \emph{holomorphic} quadric in twistor space.

This enables an ASD Maxwell field on $\R^4$ to be constructed from $Q$ via the Penrose transform~\cite{Penrose:1969ae,Eastwood:1981jy}:
\be\label{ASDMax}
\begin{split}
\tilde{\varphi}_{\alpha\beta}&=-\frac{2M}{\pi}\oint\D\lambda\,\frac{\lambda_{\alpha}\,\lambda_{\beta}}{Q\,\la \lambda|T|o]^2} \\
 &=-\frac{4\im\,M\,A_{\alpha}\,A_{\beta}}{\la B\,A\ra\,\la A|T|o]^2}=-\frac{4M}{r}\,(\iota_{\alpha}-\zeta_{-}\,o_{\alpha})\,(\iota_{\beta}-\zeta_{-}\,o_{\beta})\,,
\end{split}
\ee
where the contour integral is taken around the pole at $\la A\,\lambda\ra=0$. It is easy to see that $\nabla^{\alpha\dot\alpha}\tilde{\varphi}_{\alpha\beta}=0$, as required for an ASD Maxwell field. A theorem of Tod~\cite{Tod:1982mmp} then ensures that the metric
\be\label{ASDTN2}
\d s^2=\delta+o_{\dot\alpha}\,o_{\dot\beta}\,\tilde{\varphi}_{\alpha\beta}\,\d x^{\alpha\dot\alpha}\,\d x^{\beta\dot\beta}\,,
\ee
is vacuum, Kerr-Schild and ASD. By comparison with \eqref{ASDTN}, one immediately sees that this is precisely the ASDTN metric in Kerr-Schild form.

In other words, the holomorphic quadric $\mathcal{Q}$ in flat twistor space $\PT$ encodes -- entirely algebraically -- the ASDTN metric on $\R^4$. In contrast to the non-linear graviton, this construction is defined in twistor space, rather than the dual twistor space. Furthermore (and again unlike the non-linear graviton), it places the algebraic specialty of ASDTN on the same technical footing as the hyperk\"ahler condition through the quadric's encoding of the rank-two Killing spinor\footnote{In fact, \emph{all} vacuum, Euclidean-real, ASD slices of the Pleba\'nski-Demia\'nski~\cite{Plebanski:1976gy} family of metrics can be obtained in this way, and every (generic) holomorphic quadric in $\PT$ gives rise to such a metric~\cite{Adamo:2026obu}.}. 

Finally, $Q$ also defines a certain 2-form on $\R^4$ via a contour integral:
\be\label{symplectic}
\omega:=\frac{1}{2\pi}\oint \frac{\lambda_{\alpha}\,\lambda_{\beta}}{Q^2}\,\D\lambda\wedge\d x^{\alpha\dot\alpha}\wedge\d x^{\beta}{}_{\dot\alpha}=\im\,\frac{A_{\alpha}\,B_{\beta}\,\d x^{\alpha\dot\alpha}\wedge\d x^{\beta}{}_{\dot\alpha}}{\la A\,B\ra^3}\,.
\ee
It is easy to show that $\omega$ is non-degenerate, and $\d \omega=0$ follows from the fact that the integrand in \eqref{symplectic} is a top holomorphic form on $\R^4\times \P^1$. Thus, $\omega$ is in fact a symplectic form on $\R^4$.


\section{Constructing the Schwarzschild metric}\label{sec:Construction}

It has long been known that the Schwarzschild black hole metric can be viewed as a superposition of self-dual and anti-self-dual Taub-NUT metrics~\cite{Hawking:1976jb}, although the precise incarnations of this fact can take different forms~\cite{Kim:2024mpy}. Here we exploit the fact that \emph{both} SDTN and ASDTN can be encoded in twistor space to obtain the Schwarzschild metric using only twistorial data.


\subsection{The coincidence locus}

Let us take $\CPT$ to be the twistor space of SDTN: that is, $\R^4\times\P^1$ equipped with the complex structure defined by \eqref{SDTN-Dolb}. Now, consider the quadric $\mathcal{Q}$, defined by the zero locus of $Q=\la A\,\lambda\ra\la B\,\lambda\ra$ in $\CPT$. In $\PT$ -- the twistor space of \emph{flat} space with complex structure defined by $\dbar$ -- we saw that this is a holomorphic quadric which encodes the ASDTN metric. However, in the complex structure of $\CPT$, $\mathcal{Q}$ is definitely \emph{not} a holomorphic sub-variety.

Indeed, it is easy to see that 
\be\label{Qholo1}
\nbar Q=\frac{\la o\,\lambda\ra^2}{\la\lambda\,\hat{\lambda}\ra^2}\,\varphi_{\dot\alpha}{}^{\dot\beta}\,\hat{\lambda}_{\alpha}\,\hat{\lambda}^{\beta}\,\nabla_{\beta\dot\beta}\left(\la A\,\lambda\ra\,\la B\,\lambda\ra\right)\d x^{\alpha\dot\alpha}\,,
\ee
and for a generic point on $\mathcal{Q}$, where $\la A\lambda\ra$ or $\la B\,\lambda\ra$ vanishes, $\nbar Q$ is non-vanishing. One can then consider the \emph{coincidence locus}
\be\label{coinc}
\chi=\left\{(x,[\lambda])\in\CPT\,|\,Q=0=\nbar Q\right\}\,,
\ee
that is, the points on the quadric $\mathcal{Q}$ which \emph{do} form a holomorphic subvariety of the SDTN twistor space. Na\"ively, this coincidence locus does not appear to be very interesting: $\CPT$ is three complex-dimensional, while $\chi$ is specified by two conditions ($Q=0$ and $\nbar Q=0$), so one expects $\chi$ to have only one complex (or two real) dimensions. However, the interplay between geometric structures of SDTN (encoded in $\nbar$) and ASDTN (encoded in $Q$) mean that $\chi$ is actually only co-dimension one: the coincidence locus has complex dimension two or real dimension four.

To see this, observe that
\be\label{nbarQ1}
\nbar Q=0 \quad \Leftrightarrow \quad \alpha_{\dot\alpha}\,\alpha_{\dot\beta}\,\hat{\lambda}^{\beta}\,\nabla_{\beta}{}^{\dot\beta}\left(\la A\,\lambda\ra\,\la B\,\lambda\ra\right)=0\,.
\ee
Now, from \eqref{AB} and the fact that $T^{\alpha\dot\beta}T_{\alpha}{}^{\dot\alpha}=\frac{1}{2}\epsilon^{\dot\alpha\dot\beta}$, it follows that
\be\label{A-alpha-ident}
\alpha_{\dot\alpha}=-\sqrt{2}\,T^{\alpha}{}_{\dot\alpha}\,A_{\alpha}\,,
\ee
so that the condition $\nbar Q=0$ becomes
\be\label{nbarQ2}
T^{\delta}{}_{\dot\beta}\,\hat{\lambda}^{\beta}\,A_{\delta}\,\nabla_{\beta}{}^{\dot\beta}\left(\la A\,\lambda\ra\,\la B\,\lambda\ra\right)=0\,.
\ee
This seems a fairly unenlightening condition at this point.

However, the Killing spinor of ASDTN obeys 
\be\label{ASD-Killing}
A^{(\alpha}B^{\beta)}=\tilde{K}^{\alpha\beta}=T^{(\alpha}{}_{\dot\alpha}\,x^{\beta)\dot\alpha}\,,
\ee
so that
\be\label{nbarQ3}
\nabla_{\beta}{}^{\dot\beta}\left(\la A\,\lambda\ra\,\la B\,\lambda\ra\right)=\nabla_{\beta}{}^{\dot\beta}\left(\tilde{K}^{\rho\sigma}\,\lambda_{\rho}\,\lambda_{\sigma}\right)=T^{\rho\dot\beta}\,\lambda_{\rho}\,\lambda_{\beta}\,.
\ee
Feeding this into \eqref{nbarQ2} and using $T^{\rho\dot\beta}T^{\delta}{}_{\dot\beta}=\frac{1}{2}\epsilon^{\rho\delta}$, we obtain that the condition for $\nbar Q$ to vanish is simply
\be\label{nbarQ4}
\la A\,\lambda\ra\,\la\lambda\,\hat{\lambda}\ra=0\,,
\ee
which holds only if $\la A\,\lambda\ra$ vanishes, since $\la\lambda\,\hat{\lambda}\ra$ is strictly non-zero\footnote{Recall that in components $\lambda_{\alpha}=(\lambda_0,\lambda_1)\neq0$, so we have $\la\lambda\,\hat{\lambda}\ra=-(|\lambda_0|^2+|\lambda_1|^2)\neq0$.}.

Returning to the coincidence locus, defined by $Q=0=\nbar Q$, one discovers that $\chi$ is specified by a \emph{single} condition, namely:
\be\label{coinc2}
\chi=\left\{(x,[\lambda])\in\CPT\,|\,\la A\,\lambda\ra=0\right\}\,.
\ee
In particular, the coincidence locus corresponds to a two-dimensional, holomorphic complex subvariety of the twistor space of SDTN.


\subsection{Complex structure of the coincidence locus}

Having established that the coincidence locus corresponds to the subvariety $\la A\,\lambda\ra=0$ in $\CPT$, any homogeneous quantity can be restricted to $\chi$ simply by replacing $\lambda_{\alpha}$ with $A_{\alpha}$. In particular, this means that the Dolbeault operator \eqref{SDTN-Dolb} on $\CPT$ restricts to a Dolbeault operator on $\chi$:
\be\label{chi-Dolb}
\Dbar\vcentcolon=\nbar|_{\chi}=\frac{\hat{A}_{\alpha}\,\d x^{\alpha\dot\alpha}}{\la A\,\hat{A}\ra}\,A^{\beta}\,\frac{\partial}{\partial x^{\beta\dot\alpha}}+\frac{\la o A\ra^2}{\la A\,\hat{A}\ra^2}\,\varphi_{\dot\alpha}^{\,\,\dot\beta}\,\hat{A}_{\alpha}\,\hat{A}^{\beta}\,\d x^{\alpha\dot\alpha}\,\frac{\partial}{\partial x^{\beta\dot\beta}}\,.
\ee
It follows that this Dolbeault operator is integrable (i.e., $\Dbar^2=0$) as $\chi$ is a holomorphic subvariety of $\CPT$; this can also be established by direct calculation using the defining properties of the null SD Maxwell field $\varphi_{\dot\alpha\dot\beta}$ and the various principal spinors.

Now, observing that $\hat{o}_{\dot\alpha}=\iota_{\dot\alpha}$ and $\hat{\iota}_{\dot\alpha}=-o_{\dot\alpha}$, it follows from \eqref{AB} that
\be\label{BAhat}
\begin{split}
\la B\,\hat{A}\ra &=\sqrt{x^2+y^2}\,T^{\alpha\dot\alpha}\,T_{\alpha}{}^{\dot\beta}\left(o_{\dot\alpha}+\zeta_+\,\iota_{\dot\alpha}\right) \left(\iota_{\dot\beta}-\bar{\zeta}_{-}\,o_{\dot\beta}\right) \\
 & =\frac{\sqrt{x^2+y^2}}{2}\left(1+\zeta_+\,\bar{\zeta}_-\right)=0\,,
\end{split}
\ee
with the first equality following from $T^{\alpha\dot\alpha}T_{\alpha}{}^{\dot\beta}=\frac{1}{2}\epsilon^{\dot\alpha\dot\beta}$ and the final equality following from \eqref{spinident1}. This means that $\hat{A}_{\alpha}$ is proportional to $B_{\alpha}$, and since $\Dbar$ given by \eqref{chi-Dolb} is homogeneous in $\hat{A}_{\alpha}$, we obtain the equivalent expression
\be\label{Schwarzchild-Dolb}
\begin{split}
\Dbar &=\frac{B_{\alpha}\,\d x^{\alpha\dot\alpha}}{\la A\,B\ra}\,A^{\beta}\,\frac{\partial}{\partial x^{\beta\dot\alpha}}+\frac{\la o A\ra^2}{\la A\,B\ra^2}\,\varphi_{\dot\alpha}{}^{\dot\beta}\,B_{\alpha}\,B^{\beta}\,\d x^{\alpha\dot\alpha}\,\frac{\partial}{\partial x^{\beta\dot\beta}} \\
& = \frac{B_{\alpha}\,\d x^{\alpha\dot\alpha}}{\la A\,B\ra}\,A^{\beta}\,\frac{\partial}{\partial x^{\beta\dot\alpha}}-\frac{4\im\,M}{[\beta\,\alpha]}\,\frac{\alpha_{\dot\alpha}\,\alpha^{\dot\beta}\,B_{\alpha}\,B^{\beta}}{\la A\,B\ra^2}\,\d x^{\alpha\dot\alpha}\,\frac{\partial}{\partial x^{\beta\dot\beta}}\,.
\end{split}
\ee
This form of the Dolbeault operator on $\chi$ will be particularly useful going forward.

\medskip

At this point, we wish to associate an integrable complex structure on $\chi$ to its induced Dolbeault operator $\Dbar$. Crucially, this complex structure should be constructed \emph{only} from the data in $\Dbar$ (that is, without having to refer to a conjugate Dolbeault operator); to do this, we follow the approach of~\cite{Araneda:2022xii}.

The action of $\Dbar$ maps $\Omega^0_{\chi}\to\Omega^{0,1}_{\chi}$, where $\Omega^0_{\chi}$ is the space of functions on the coincidence locus and $\Omega^{0,1}_{\chi}$ is the space of 1-forms on $\chi$ which are $(0,1)$-forms in the complex structure we wish to associate to $\Dbar$. Let $T^{0,1}_{\chi}$ denote the dual space of vector fields. From \eqref{Schwarzchild-Dolb}, it is easy to identify a basis of such $(0,1)$-forms and vectors:
\be\label{0,1-forms}
\Omega^{0,1}_{\chi}=\mathrm{span}\,\left\{\frac{B_{\alpha}\,\d x^{\alpha\dot\alpha}}{\la A\,B\ra}\right\}:=\mathrm{span}\left\{\tilde{\theta}^{\dot\alpha}\right\}\,,
\ee
\be\label{0,1-vecs}
T^{0,1}_{\chi}=\mathrm{span}\,    \left\{A^{\beta}\,\frac{\partial}{\partial x^{\beta\dot\alpha}}-\frac{\la o A\ra^2}{\la A\,B\ra}\,\varphi_{\dot\alpha}^{\,\,\dot\beta}\,B^{\beta}\,\frac{\partial}{\partial x^{\beta\dot\beta}}\right\} :=\mathrm{span}\left\{\tilde{T}_{\dot\alpha}\right\}\,,
\ee
respectively. These bases are normalized so that $\tilde{T}_{\dot\alpha} \lrcorner\tilde{\theta}^{\dot\beta}=\delta_{\dot\alpha}^{\dot\beta}$, where $\lrcorner$ denoted the interior product between a vector and a 1-form.

Bases of the spaces of $(1,0)$-forms and $(1,0)$-vectors -- denoted by $\Omega^{1,0}_{\chi}$ and $T^{1,0}_{\chi}$, respectively -- can then be obtained by demanding that their inner product with the $(0,1)$-vectors or $(0,1)$-forms vanishes. This leads to
\be\label{1,0-forms}
\Omega^{1,0}_{\chi}=\mathrm{span}\,\left\{A_{\alpha}\,\d x^{\alpha\dot\alpha}-\frac{\la o\,A\ra^2}{\la A\,B\ra}\,\varphi^{\dot\alpha}{}_{\dot\beta}\,B_{\beta}\,\d x^{\beta\dot\beta}\right\}:=\mathrm{span}\left\{\theta^{\dot\alpha}\right\}\,,
\ee
\be\label{1,0-vecs}
T^{1,0}_{\chi}=\mathrm{span}\,\left\{\frac{B^{\beta}}{\la B\,A\ra}\,\frac{\partial}{\partial x^{\beta\dot\alpha}}\right\}:=\mathrm{span}\left\{T_{\dot\alpha}\right\}\,.
\ee
It is easy to see that these obey $T_{\dot\alpha}\lrcorner\tilde{\theta}^{\dot\beta}=0=\tilde{T}_{\dot\alpha}\lrcorner\theta^{\dot\beta}$ as well as $T_{\dot\alpha}\lrcorner\theta^{\dot\beta}=\delta^{\dot\beta}_{\dot\alpha}$. 

The complex structure corresponding to this decomposition of $T_{\chi}=T^{1,0}_{\chi}\oplus T^{0,1}_{\chi}$ is then
\be\label{J}
\begin{split}
J&=\im\left(T_{\dot\alpha}\otimes\theta^{\dot\alpha}-\tilde{T}_{\dot\alpha}\otimes\tilde{\theta}^{\dot\alpha}\right) \\
&=\im\left(\delta^{\dot\beta}_{\dot\alpha}\,\frac{B_{\alpha}\,A^{\beta} +A_{\alpha}\,B^{\beta}}{\la A\,B\ra}+2\frac{\la o A\ra^2}{\la A\,B\ra^2}\,\varphi_{\dot\alpha}^{\,\,\dot \beta}\,B_{\alpha}\,B^{\beta}\right)\frac{\partial}{\partial x^{\beta\dot\beta}}\otimes \d x^{\alpha\dot\alpha} \\
&=\im\left(\delta^{\dot\beta}_{\dot\alpha}\,\frac{B_{\alpha}\,A^{\beta} +A_{\alpha}\,B^{\beta}}{\la A\,B\ra}-8\im\,M\frac{\alpha_{\dot\alpha}\,\alpha^{\dot\beta}\,B_{\alpha}\,B^{\beta}}{[\beta\,\alpha]\,\la A\,B\ra^2}\right)\frac{\partial}{\partial x^{\beta\dot\beta}}\otimes \d x^{\alpha\dot\alpha}\,.
\end{split}
\ee
It is easy to confirm that $T^{1,0}_{\chi}$ and $T^{0,1}_{\chi}$ are the $\pm\im$ eigenspaces of $J$, respectively.


\subsection{K\"ahler structure of the coincidence locus \& the Schwarzschild metric}

Having obtained the complex structure \eqref{J} on the coincidence locus induced by the complex structure on $\CPT$, we now observe that $\chi$ also comes equipped with a natural symplectic form -- namely, the $\omega$ defined by the quadric $Q$ itself in \eqref{symplectic}. Remarkably, a simple calculation shows that
\be\label{compatible1}
\omega(J,J)=\omega\,,
\ee
so the complex and symplectic structures on $\chi$ are compatible\footnote{This compatibility relies on the fact that $\hat{A}_{\alpha}\propto B_{\alpha}$, following from \eqref{BAhat}.}. Thus, $\chi$ comes equipped with a K\"ahler structure defined by the data $(\omega,J)$. 

The corresponding K\"ahler metric is then obtained as $g_{\mathrm{K}}:=\omega(\cdot,J)$, which is guaranteed to be symmetric thanks to the compatibility between the symplectic form and complex structure. This K\"ahler metric is given explicitly by
\be\label{Kahler-met}
\begin{split}
g_{\mathrm{K}}&=-\left(\delta^{\dot\gamma}_{\dot\alpha}\,\frac{B_{\alpha}\,A^{\gamma}+A_{\alpha}\,B^{\gamma}}{\la A\,B\ra}+2\,\frac{\la o\, A\ra^2}{\la A\,B\ra^2}\,\varphi_{\dot\alpha}{}^{\dot \gamma}\,B_{\alpha}\,B^{\gamma}\right) \frac{A_{(\gamma}\, B_{\beta)}\, \epsilon_{\dot \gamma \dot \beta}}{\la A B \ra^3}\,\d x^{\alpha\dot\alpha}\,\d x^{\beta\dot\beta} \\
&=\frac{-1}{\la A B \ra^2}\left(\epsilon_{\alpha \beta}\,\epsilon_{\dot \alpha \dot \beta}+\frac{\la o\, A \ra^2}{\la A B \ra^2}\,\varphi_{\dot\alpha\dot\beta}\,B_{\alpha}\,B_{\beta}\right)\d x^{\alpha\dot\alpha}\,\d x^{\beta\dot\beta} \\
&= \frac{-1}{\la A\,B\ra^2}\left[\delta- \frac{4\im\,M}{[\beta\,\alpha]}\left(\frac{\alpha_{\dot \alpha}\,B_{\alpha}}{\la A\, B \ra}\,\d x^{\alpha\dot\alpha}\right)^2\right]\,.
\end{split}
\ee
Thus, we see that the coincidence locus inherits a K\"ahler metric which is conformally related to a Kerr-Schild metric.

Already this is a surprising result: using only holomorphic data on twistor space, we have obtained a 4-dimensional metric which is K\"ahler and non-self-dual. However, more remarkable is the fact that upon using the identities \eqref{spinident1} and \eqref{AB}, $g_{\mathrm{K}}$ can be expressed as
\be\label{confKahler1}
g_{\mathrm{K}}=\frac{1}{r^2}\,g_{\mathrm{S}}\,,
\ee
where
\be\label{Schwarzschild}
g_{\mathrm{S}}:=\delta+\frac{4\,M}{r}\,\ell_{\alpha\dot\alpha}\,\ell_{\beta\dot\beta}\,\d x^{\alpha\dot\alpha}\,\d x^{\beta\dot\beta}\,, \qquad \quad \ell_{\alpha\dot\alpha}:=-\im\,\frac{B_{\alpha}\,\alpha_{\dot\alpha}}{r}\,,
\ee
is the Schwarzschild metric in Kerr-Schild coordinates~\cite{Kerr:1965vyg,Debney:1969zz}. In particular, we have recovered the Schwarzschild metric using holomorphic data defined only on twistor space.

This result can be repackaged as the following:
\begin{thm}\label{MainThm}
Let $(\CPT,\nbar)$ be the twistor space of self-dual Taub-NUT in Kerr-Schild gauge, and $\chi\subset\CPT$ be the coincidence locus defined by $Q=0=\nbar Q$, where $Q=\la A\,\lambda\ra\,\la B\,\lambda\ra$ is defined by the principal spinors of the anti-self-dual Taub-NUT metric in Kerr-Schild gauge. Then $\dim_{\C}\chi=2$ and $\chi$ inherits a K\"ahler metric conformal to the Schwarzschild metric in Kerr-Schild gauge.
\end{thm}

\medskip

Note that the Schwarzschild metric resulting from this construction can be defined in either Euclidean \emph{or} Lorentzian signature. Of course, the inputs for this construction were SDTN and ASDTN in Euclidean signature, but in fact the only reality conditions required at each stage in the construction were that $x,y,z\in\R$. Indeed, these are the reality conditions underpinning crucial relations like \eqref{spinident1} and \eqref{BAhat}; the stationary nature of the metrics means that reality conditions on $\tau$ are never required. Consequently, one can either take $\tau\in\R$ to obtain the Euclidean Schwarzschild metric in \eqref{Schwarzschild}, or $\tau=\im\,t$ with $t\in\R$ to obtain the Lorentzian Schwarzschild black hole. The only choice of signature which is \emph{not} allowed in the construction is split, or ultrahyperbolic, signature, which would require one of $x,y,z$ to be purely imaginary.

Finally, a remark on topology. Both the Riemannian Schwarzschild instanton and the Lorentzian Schwarzschild black hole have topology $\R^2\times S^2$ (with different coordinates playing different roles depending on the signature).  One might wonder where this topology is encoded in our construction, which started from a smooth metric on $\R^4$ (namely, the Euclidean SDTN metric). However, the construction here is entirely local; that is, \eqref{Schwarzschild} should be viewed as a metric defined on a local coordinate patch which is indistinguishable from an open subset of $\R^4$. Recovering the global topology follows by fixing reality conditions (i.e., choosing Euclidean or Lorentzian signature) and then considering the underlying manifold topology under which the resulting real metric can be globally extended.


\subsection{Einstein condition}
\label{sec:Einstein}

One of the key features of twistor theory is that field equations on spacetime are solved implicitly in terms of analytic data. This is certainly also the case in our construction, where the Schwarzschild metric is obtained without any explicit reference to the vacuum Einstein equations. The fact that the metric $g_{\mathrm{S}}$ in \eqref{Schwarzschild} solves the vacuum Einstein equations follows immediately upon realising that it is a known exact solution to these equations. Nevertheless, one might wonder if there is a canonical reason to identify the conformal factor of $r^{-1}$ in the relationship between $g_{\mathrm{K}}$ and $g_{\mathrm{S}}$.

In fact, there is a clear reason for doing so based on a remarkable theorem of Derdzinski, which states\footnote{Derdzinski's original statement -- corresponding to Proposition 4 in~\cite{Derdzinski:1983} -- contains more than what we are using here.}:
\begin{thm}[Derdzinski~\cite{Derdzinski:1983}]
Let $(g_{\mathrm{K}},J,\omega)$ be a 4-dimensional K\"ahler metric with non-constant scalar curvature $R_{g_{\mathrm{K}}}$. Suppose that $g_{\mathrm{K}}$ is extremal K\"ahler and non-self-dual -- that is, $V=\omega^{-1}(R_{g_{\mathrm{K}}},\cdot)$ is a holomorphic Killing vector. Then if
\be\label{Der-eq}
R_{g_\mathrm{K}}^{3}+6R_{g_\mathrm{K}}\,\Delta R_{g_\mathrm{K}}-12\,|\nabla R_{g_\mathrm{K}}|^2=\mathrm{constant}\,,
\ee
for $\Delta:=g_{K}^{ab}\,\nabla_{a}\nabla_{b}$, the metric
\be\label{Der-Ein-met}
\tilde{g}:=\frac{1}{R^2_{g_{\mathrm{K}}}}\,g_{\mathrm{K}}\,,
\ee
is Einstein.
\end{thm}
In other words, this theorem gives a criterion for an extremal K\"ahler metric to be conformal to an Einstein metric, with conformal factor given by the inverse of the scalar curvature.

In fact, this condition can be recast in a slightly more transparent form. Let $\Omega:=R_{g_{\mathrm{K}}}^{-1}$ and $R_{\tilde{g}}$ be the scalar curvature of the metric \eqref{Der-Ein-met}. Then using the well-known formula for the transformation of the Ricci scalar under a conformal transformation, \eqref{Der-eq} becomes
\be\label{Der-eq2}
\Omega^{-2}\left(R_{g_{\mathrm{K}}}-6\Omega^{-1}\,\Delta\Omega\right)=R_{\tilde{g}}=\mathrm{constant}\,,
\ee
as expected for an Einstein metric. So, for the conformal metric $\tilde{g}$ to be a vacuum metric, the constant on the right-hand side of \eqref{Der-eq} must be zero.

\medskip

Clearly in our case, we wish to take $g_{\mathrm{K}}$ to be the K\"ahler metric \eqref{confKahler1} on the coincidence locus. First, the scalar curvature of this metric can be computed directly:
\be\label{gKScalar}
R_{g_{\mathrm{K}}}=\frac{24\,M}{r}\,,
\ee
which is non-constant. Now, with the Poisson bi-vector dual to the K\"ahler form given by
\be\label{inverseKahler}
\omega^{-1}=-\im\,\la A\,B\ra \,A^{\alpha}\,B^{\beta}\,\frac{\partial}{\partial x^{\alpha}{}_{\dot\alpha}}\wedge\frac{\partial}{\partial x^{\beta\dot\alpha}}\,,
\ee
it follows that
\be\label{gKKilling}
V=24M\,\partial_{\tau}\,,
\ee
which is indeed a holomorphic Killing vector of $g_{\mathrm{K}}$. Thus, all the conditions for applying Derdzinski's theorem are met.

Now, an explicit computation gives
\be\label{Derd-eq3}
\Delta\Omega=\frac{1}{24M\,\sqrt{\det g_{\mathrm{K}}}}\,\partial_{a}\left(\sqrt{\det g_{\mathrm{K}}}\,g^{ab}_{\mathrm{K}}\,\partial_{b}r\right)=\frac{1}{6}\,,
\ee
so that
\be\label{Derd-eq4}
1-6\,\Delta\Omega=0 \quad \Rightarrow \quad R_{\tilde{g}}=0\,.
\ee
Namely, Derdzinski's theorem implies that the metric $\tilde{g}$ automatically solves the vacuum Einstein equations. Of course, $\tilde{g}=(24M)^{-2}g_{\mathrm{S}}$, so the resulting vacuum Einstein metric \emph{is} the Schwarzschild metric after a suitable (constant) rescaling.

What one learns from this exercise is that the identification of $g_{\mathrm{K}}=r^{-2}\,g_{\mathrm{S}}$ is no accident: $r^{-1}$ is \emph{the} conformal factor under which the K\"ahler metric $g_{\mathrm{K}}$ is related to a vacuum Einstein metric.


\section{Discussion}
\label{sec:Disc}

In this paper, we provided a resolution of the googly problem in the specific case of the Schwarzschild metric. This construction took as inputs an array of ideas: the relationship between K\"ahler geometry, the type D condition and quadric hypersurfaces; the picture of Schwarzschild as a superposition of SD and ASD Taub-NUT metrics; and the availability of Kerr-Schild coordinates for all three metrics. The result is obtained by considering the coincidence locus in the twistor space of SDTN; that is, a holomorphic subvariety of the twistor space which is associated to the ASDTN geometry.

This is the first example of a solution to the googly problem, albeit for a specific metric. Note that it has long been known that twistor theory can generate toric, Ricci-flat metrics (such as Schwarzschild) indirectly through stationary, axisymmetric SL$(2,\C)$ SD Yang-Mills fields on flat space~\cite{Witten:1979tv,Ward:1982bf,Dunajski:2024myp}. However, in contrast to this procedure, our construction is manifestly gravitational, is set in a curved (rather than flat) twistor space, and explicitly tracks the SD and ASD degrees of freedom in the resulting metric. 

Of course, the Schwarzschild metric is one of the best-known exact solutions to the vacuum Einstein equations; one might ask what is gained by obtaining such a well-studied solution by the apparently Byzantine methods of twistor theory. The key advantage here is the fact that the coincidence locus lives within the twistor space of SDTN. This raises the possibility that exact calculations on SDTN using twistor theory, including closed-form wavefunctions and exact scattering amplitudes~\cite{Adamo:2023fbj,Guevara:2023wlr,Araneda:2024cqu,Guevara:2024edh,Adamo:2025fqt,Doran:2026bng}, could induce structures on the coincidence locus which encode information about scattering on Schwarzschild. We hope to explore such a twistor approach to black hole perturbation theory in future work.

Finally, it is natural to ask if other exact solutions of the Einstein equations can be obtained in this way. In particular, the Kerr metric seems a natural candidate to admit a googly description which generalizes the one given here for Schwarzschild, and the construction in~\cite{Araneda:2022xii} shows that it can be described as a deformed twistor quadric. Other black hole metrics may then correspond to appropriate subvarieties of twistor spaces, generalizing the coincidence locus studied here. The field equations will be guaranteed to be satisfied by the argument given in section~\ref{sec:Einstein}, and the deformation of the complex structure will be captured by new SU$(\infty)$ Toda data, generalizing the deformations of hyperk\"ahler structures described in~\cite{Adamo:2026obu}.

\acknowledgments

We thank Joon-hwi Kim for interesting conversations. The authors are supported by a Royal Society University Research Fellowship (TA), the Simons Collaboration on Celestial Holography CH-00001550-11 (TA \& SS), the ERC Consolidator/UKRI Frontier grant TwistorQFT EP/Z000157/1 (TA \& BA), the STFC consolidated grant ST/X000494/1 (TA), and the Gordon and Betty Moore Foundation and the John Templeton Foundation via the Black Hole
Initiative (AS). The research of BA was also supported by the Simons Foundation grant SFI-MPS-T-Institutes-00010825, and by the Polish State Treasury funds as part of a task commissioned by the Minister of Science and Higher Education under the project Organization of the Simons Semesters at the Banach Center- New Energies in 20262028 (MNiSW/2025/DAP/491).

\bibliographystyle{JHEP}
\bibliography{kerr}

\end{document}